\documentclass[%
 reprint,
 amsmath,amssymb,
 aps,
]{revtex4-2}

\usepackage{graphicx}% Include figure files
\usepackage{dcolumn}% Align table columns on decimal point
\usepackage{bm}% bold math
\usepackage{physics}
\usepackage{siunitx}
\usepackage{dsfont}
\usepackage{layouts}
\usepackage[colorlinks=true]{hyperref}
\newcommand{\me}{\text{e}}
\newcommand{\imi}{\text{i}}

\newcommand{\Qprop}{Q{\textsc{prop}}}

\newcommand{\figref}[1]{Fig.~\ref{#1}}
\begin{document}

\title{Advanced momentum sampling and Maslov phases for a precise semiclassical model of strong-field ionization}

\author{Mads Brøndum Carlsen}
\author{Emil Hansen}
\author{Lars Bojer Madsen}
\author{Andrew Stephen Maxwell}
\email{andrew.maxwell@phys.au.dk} 
\affiliation{Department of Physics and Astronomy, Aarhus University, 8000 Aarhus C, Denmark}

\date{\today}

\begin{abstract}
Recollision processes are fundamental to strong-field physics and attoscience, thus models connecting recolliding trajectories to quantum amplitudes are a crucial part in furthering understanding of these processes.
We report developments in the semiclassical path-integral-based Coulomb quantum-orbit strong-field approximation model for strong-field ionization by including an additional phase known as Maslov's phase and implementing a new solution strategy via Monte-Carlo-style sampling of the initial momenta. In doing so, we obtain exceptional agreement with solutions to the time-dependent Schrödinger equation for hydrogen, helium, and argon.
%something that until recently was beyond reach with semiclassical models.
We provide an in-depth analysis of the resulting photoelectron momentum distributions for these targets, facilitated by the quantum-orbits arising from the solutions to the saddle-point equations. The analysis yields a new class of rescattered trajectories that includes the well-known laser-driven long and short trajectories, along with novel Coulomb-driven rescattered trajectories.
By virtue of the precision of the model, it opens the door to detailed investigations of a plethora of strong-field phenomena such as photoelectron holography, laser-induced electron diffraction and high-order above threshold ionization.
\end{abstract}

\maketitle

\section{Introduction}

%attosecond physics and strong-field physics---importance of recollision
Strong-field physics is the study of intense and short laser fields interacting with matter, which has enabled the measurement and manipulation of electrons on an attosecond timescale ($10^{-18}$~s). The strong-field process high-harmonic generation (HHG) \cite{corkum_plasma_1993,lewenstein_theory_1994} has enabled the production of ultra-short attosecond laser pulses \cite{paul_observation_2001,hentschel_attosecond_2001} and given rise to the field of attosecond physics \cite{krausz_attosecond_2009, salieres_study_1999, lewenstein_principles_2009, ciappina_attosecond_2017}. The process of above-threshold ionization (ATI) \cite{agostini_freefree_1979} (specifically strong-field ionization \cite{popruzhenko_keldysh_2014}) has been used in ultrafast imaging procedures, e.g., laser-induced electron diffraction (LIED) \cite{zuo_laserinduced_1996,niikura_sublasercycle_2002,amini_chapter_2020a,sanchez_molecular_2021,giovannini_new_2023}, and photoelectron holography \cite{huismans_timeresolved_2011,hickstein_direct_2012,faria_it_2020}.
All of these processes and applications are underpinned by recollision or recombination of the photoelectron. As such, all the above advancements and applications were enabled by breakthroughs in understanding and modelling the photoelectron's return.

%Early models sucesses and inadquacies
%Idea of laser driven recollision
The strong-field approximation (SFA) \cite{keldysh_ionization_1965,faisal_multiple_1973,reiss_effect_1980} (see Ref.~\cite{amini_symphony_2019} for a historical overview) and the three-step model \cite{corkum_plasma_1993}, enabled significant progress in the description of HHG \cite{lewenstein_theory_1994} and ATI \cite{paulus_rescattering_1994,lewenstein_rings_1995}, through the understanding of laser-driven returns of the photoelectron. 
Recollision models have enabled simple quantification of the electron's maximum return energy \cite{corkum_plasma_1993} ($3.17 U_p$) in terms of the ponderomotive energy ($U_p$), the quiver energy of a free electron in a laser field. This in turn led to a simple equation for the HHG photon energy cut-off \cite{lewenstein_theory_1994} ($I_p+3.17 U_p$), $I_p$ being the ionization potential of the target, as well as the cut-off for the direct and rescattered photoelectron energy \cite{paulus_rescattering_1994} ($2 U_p$ and $10 U_p$, respectively).
It was found that HHG and high-energy ATI energy spectra could be well-described by two types of trajectories undergoing a laser-driven recollision with shorter or longer time before returning \cite{lewenstein_phase_1995}, known as the short and long trajectories.
Such insight led directly to considerable experimental developments in HHG, see, e.g., Refs.~\cite{schiessl_quantum_2007,zair_quantum_2008,brugnera_trajectory_2011}. In the case of ATI, recollision also plays a crucial role in LIED \cite{zuo_laserinduced_1996,niikura_sublasercycle_2002}, where laser-driven photoelectron recollision is used to image molecules, or photoelectron holography \cite{huismans_timeresolved_2011,hickstein_direct_2012}, where recolliding wavepackets interfere with those that do not recollide. Considerable success of the such recollision models can be attributed to their simplicity, e.g., the Coulomb-field was neglected in the continuum, often allowing an analytic description.

%Newer more accurate models including the Coulomb potential
%Current draw back of these models
Experimental breakthroughs \cite{rudenko_resonant_2004,maharjan_wavelength_2006} measuring accurate two- or even three-dimensional momentum (see, e.g., Ref.~\cite{gopal_threedimensional_2009}) distributions for the photoelectron in strong-field ionization (instead of photoelectron energy spectra), revealed that models that neglect the Coulomb field during continuum propagation could not reproduce many key features for linearly polarized fields. Following this, a number of new models were developed that could properly account for the effect of the Coulomb potential for the photoelectron in the continuum, see, e.g., \cite{popruzhenko_coulombcorrected_2008,smirnova_analytical_2008,yan_lowenergy_2010,torlina_timedependent_2012,li_classicalquantum_2014,lai_influence_2015,shvetsov-shilovski_semiclassical_2016,maxwell_coulomb-corrected_2017}, and Ref.~\cite{faria_it_2020} for a review.
The majority of these models took a semiclassical approach, applying different approximations and expansion upon Feynman path integrals. With these approaches, a qualitative description of strong-field ionization was possible, while retaining some descriptive power of recolliding trajectories.

%State of the art, Lein Maslov phases
%CQSFA forward paper
Semiclassical models describing strong-field ionization fall into two categories \cite{faria_it_2020,rodriguez_forward_2023}, `forward' methods that propagate many trajectories, binning the final momenta, or `inverse' methods that find the specific trajectories that lead to each final momentum point. The benefit of forward methods is that it is easy to implement and generalize, however, it requires a huge number of trajectories to converge, which makes analysis of trajectories particularly difficult, as well as being numerically challenging. The inverse method provides much easier and clean trajectory-based analysis, while also being much faster to compute. However, the downside is that the inverse method is more difficult to implement and has been less generalizable.
The Coulomb quantum-orbit strong-field approximation (CQSFA) \cite{lai_influence_2015,maxwell_coulomb-corrected_2017,maxwell_analytic_2018,maxwell_coulomb-free_2018} is the primary inverse method that has provided a powerful trajectory-based description of strong-field ionization \cite{kang_holographic_2020,maxwell_spirallike_2020,werby_dissecting_2021,werby_probing_2022}. However, until recently, this method was only capable of qualitative agreement with solutions of the time-dependent Shr\"odinger equation (TDSE). The primary reasons were that important recolliding trajectories were missing, it lacked crucial phases known as the Maslov phase, and key features like pulse envelopes and solving in full three-dimensions instead of two-dimensions were missing. The Maslov phases had been neglected in all semiclassical treatments of strong-field ionization, until its recent inclusion in Ref.~\cite{brennecke_gouys_2020}. The Maslov phase is a geometric winding number \cite{tsue_maslov_1992} that has an analogy with the Gouy phase in optics \cite{petersen_unifying_2014}.

%What we are including in this work, and implications for laser driven recollision
In this work, we address all of these issues, presenting a range of developments in the CQSFA that enables a high-level of quantitative agreement with a TDSE solution. Namely, using a similar methodology to \cite{brennecke_gouys_2020}, we implement the Maslov phases, using an adaptive Monte-Carlo-style momentum sampling methods we capture all valid trajectories, we implement a sin-squared pulse envelope, and provide solutions in three dimensions. To benchmark these changes, we compare the CQSFA to TDSE solutions for three different atomic targets and obtain exceptional agreement. Then, using a trajectory-based analysis, we identify a new class of rescattered solutions that capture all important recollision dynamics. This includes the well-known long and short laser-driven recolliding trajectories, but also trajectories whose return is dominated by the Coulomb potential, and cases that fit in-between these two extremes. We directly link these trajectories to the final momentum distributions and initial momentum, and use this to devise further improvements to the sampling algorithm.

This paper is organized as follows. In Sec.~\ref{sec:theory}, we present the theory for the CQSFA, giving a brief overview in Sec.~\ref{sec:theory:CQSFA}, and proving more detail for the newest developments, the calculation of Maslov phases (Sec.~\ref{sec:theory:Maslov}), the momentum sampling algorithm in Sec.~\ref{sec:theory:Inversion}, while in Sec.~\ref{sec:theory:orbits} we define the CQSFA orbits. In Sec.~\ref{sec:Results}, we present our main results, providing a comparison of the CQSFA with the TDSE for three different atomic targets, and providing an in-depth analysis of the ionization times, trajectories, and their contribution to the photoelectron momentum distributions. Finally, in Sec.~\ref{sec:conclusion}, we present our conclusions.

\section{Theory \label{sec:theory}}
In this section, a short deviation of the CQSFA is presented in order to highlight some small deviations from the original formulation \cite{lai_influence_2015,maxwell_coulomb-corrected_2017}. The following section provides some details on the calculation of the Maslov phase \cite{brennecke_gouys_2020}. Throughout, we will work in atomic units (a.u.) unless otherwise stated.

\subsection{The CQSFA \label{sec:theory:CQSFA}}
Our starting point is the $S$-matrix amplitude \cite{lai_influence_2015} 
\begin{equation} \label{S_mat_dyson}
    M_{\bm p_f} = -i \lim_{t \rightarrow \infty}\int^t_{-\infty} dt' \mel*{\psi_{\bm p_f}(t)}{U(t,t')H_I(t')}{\psi_0(t')},
\end{equation}
describing the transition from initial system state $\ket{\psi_0}$, to a final scattering state $\ket*{\psi_{\bm p_f}}$ with asymptotic momentum $\bm p_f$. The Hamiltonian has been separated into a field-free part $H_0$ and time-dependent interaction part $H_I(t)$
\begin{equation}
    H(t) = H_0 + H_I(t),
\end{equation}
and $U(t,t_0)$ is the time-evolution operator for the entire system. We consider atomic systems within the single-active electron approximation (SAE), such that the field-free Hamiltonian takes the form
\begin{equation}
    H_0 = \frac{1}{2}\bm p^2 + V(\bm r),
\end{equation}
where $\bm p$ is the canonical momentum of the electron and $V(\bm r)$ is the effective SAE potential. For all targets considered in this text, the potential is of the form
\begin{equation}\label{potential}
    V(\bm r) = -\frac{Z + f(r)}{r}
\end{equation}
where $Z$ is the charge of the parent ion and
\begin{equation}\label{f_function}
    f(r) = a_1\me^{-a_2 r} + a_3 r\me^{-a_4 r} + a_5\me^{-a_6 r},
\end{equation}
with coefficients from Ref.~\cite{tong_empirical_2005}.

We work exclusively within the electric dipole approximation, for which the interaction Hamiltonian in the length gauge is given by
\begin{equation}
    H_I(t) = \bm E(t)\cdot\bm r,
\end{equation}
where the electric field is related to the vector potential $\bm A(t)$ as $\bm E(t) = -  \partial_t\bm A(t)$.

To proceed, we approximate the final continuum eigenstate as a field-dressed plane wave $\ket*{\psi_{\bm p_f}(t)} \simeq \ket{\bm p_f + \bm A(t)} \equiv \ket{\tilde{\bm p}_f(t)}$, and insert the identity
\begin{equation}
    \mathds{1} = \int d\tilde{\bm p}_0(t') \ket{\tilde{\bm p}_0(t')}\bra{\tilde{\bm p}_0(t')}
\end{equation}
in Eq. \eqref{S_mat_dyson}
\begin{equation} \label{CQSFA_init_trans_amp}
\begin{aligned}
    M_{\bm p_f} ={}& -i \lim_{t \rightarrow \infty}\int_{-\infty}^t \dd{t'} \int \dd[3]{\tilde{\bm p}}_0\!(t') e^{iI_p t'} d(\tilde{\bm p}_0,t')\\
    &\times \mel{\tilde{\bm p}_f(t)}{U(t,t')}{\tilde{\bm p}_0(t')},
\end{aligned}
\end{equation}
with $I_p$ the ionization potential of the target, and the ionization matrix element
\begin{equation} \label{SFA_mel}
    d(\tilde{\bm p}_0,t) = \bra{\tilde{\bm p}_0(t)}H_I(t)\ket{\psi_0},
\end{equation}
where $\ket{\psi_0}$ is the time-independent initial state. In this way, we capture the time-evolution following ionization using the propagator $\mel{\tilde{\bm p}_f(t)}{U(t,t')}{\tilde{\bm p}_0(t')}$ from an initial dressed momentum state $\ket{\tilde{\bm p}_0(t')}$ to a final dressed momentum state $\ket{\tilde{\bm p}_f(t)}$. The semiclassical nature of the CQSFA stems from approximating this propagator by a semiclassical expression.

In the CQSFA, the propagator in Eq. \eqref{CQSFA_init_trans_amp} is written in terms of path integrals, to which semiclassical approximations can be applied. This is done by time slicing, where $U(t,t')$ is split into $N+1$ operators using its composition property, each propagating a time $\Delta t$. Between these operators the identity is resolved in dressed momentum states and position states, allowing one to write \cite{kleinert}
\begin{equation*}
    \mel{\tilde{\bm p_f}(t)}{U(t,t')}{\tilde{\bm p}_0(t')} \!= \!\prod_{n=1}^{N} 
    \!\qty[\int \!\dd[3] \tilde{\bm p}_n]
    \prod_{n=1}^{N+1} 
    \!\qty[\int \!\frac{\dd[3] \bm r_n}{(2\pi)^3}]
    e^{i \mathcal{A}_N},
\end{equation*}
where 
\begin{equation} \label{pp_N_action}
    \mathcal{A}_N = -\sum_{n=1}^{N+1} \qty[\bm r_n \cdot \qty(\tilde{\bm p}_n - \tilde{\bm p}_{n-1}) + \Delta t H(\tilde{\bm p}_n, \bm r_n, t_n)].
\end{equation}
The limit $N \rightarrow \infty$, is now taken, in order to turn the product of integrals into path integrals. 

The above formulation used the transition matrix element in a momentum representation, propagating from an initial momentum $\tilde{\bm p}_0$ to a final $\tilde{\bm p}_f$. It is also possible to work in a `mixed' representation, where propagation instead is from an initial position $\bm r_0$ to a final momentum \cite{shvetsov-shilovski_semiclassical_2016}. The two forms of the matrix elements can be related as 
\begin{equation}
    \mel{\tilde{\bm p}_f}{U(t,t')}{\bm r_0} = \int \dd[3] \tilde{\bm p}_0 e^{-i\tilde{\bm p}_0 \cdot \bm r_0} \mel{\tilde{\bm p}_f}{U(t,t')}{\tilde{\bm p}_0},
\end{equation}
which, together with Eq. (\ref{pp_N_action}), shows that the `sliced' action in a mixed representation includes an additional boundary term
\begin{equation} \label{px_N_action}
    \mathcal{S}_N = \mathcal{A}_N- \tilde{\bm p}_0 \cdot \bm r_0.
\end{equation}
Returning to Eq. (\ref{CQSFA_init_trans_amp}) we see that if we pull out an exponential dependence of $d$, Eq. (\ref{SFA_mel}), 
\begin{equation}\label{the_d_tilde}
     d(\tilde{\bm p}_0,t) = e^{-i\tilde{\bm p}_0 \cdot \bm r_0} \tilde{d}(\tilde{\bm p}_0,\bm r_0,t),
\end{equation}
we can write the transition amplitude in the mixed representation
\begin{equation}
    \begin{split}
    M_{\bm p_f} =&{} -i \lim_{t \rightarrow \infty} \int_{-\infty}^t \dd t' \prod_{n=0}^{N} \qty[\int \dd[3] \tilde{\bm p}_n] \prod_{n=1}^{N+1} \qty[\int \frac{\dd[3] \bm r_n}{(2\pi)^3}] \\
    &\times e^{i \mathcal{S}_N} e^{iI_p t'} \tilde{d}(\tilde{\bm p}_0, \bm r_0, t'),
    \end{split}
\end{equation}
now including the $\tilde{\bm p}_0$ integral in the product. In the limit $N\rightarrow \infty$, we obtain the path integrals 
\begin{equation}
    M_{\bm p_f} = -i \lim_{t \rightarrow \infty} \int_{-\infty}^t \dd t' 
    \int^{\mathbf{p}_f} \mathcal{D}\tilde{\bm p} \int_{\mathbf{r}_0}\frac{\mathcal{D}\bm r}{(2\pi)^3} e^{iS} \tilde{d}(\tilde{\bm p}_0, \bm r_0, t'),
\end{equation}
with the mixed representation action 
\begin{equation} \label{action}
    S(\mathbf{p}, \mathbf{r}, t') = I_pt' -\int_{t'}^{\infty} \dd \tau \qty[\dot{\bm p} \cdot \bm r(\tau) + H(\bm{p}, \bm r, \tau)] - \tilde{\bm p}_0 \cdot \bm r_0
\end{equation}
and Hamiltonian 
\begin{equation}
    H(\bm p, \bm r, \tau) = \frac{1}{2}[\bm p(\tau) + \bm A(\tau)]^2 + V[\bm r(\tau)].\
\end{equation}
The fact that we have an action corresponding to a mixed representation is the key point of this section. The effect of this is that when we apply the saddle-point approximation (SPA) to the path integrals, in order to obtain a semiclassical description, we obtain a stability matrix of the form $\partial \bm p(t) / \partial \bm p_0$, instead of the $\partial \bm p(t) / \partial \bm r_0$ that is obtained in a momentum representation \cite{miller_classical-limit_1974, shvetsov-shilovski_semiclassical_2016}. 

Next, we perform the SPA on the time and path integrals \cite{kleinert, lai_influence_2015}, thereby obtaining the desired transition amplitude
\begin{equation} \label{CQSFA_amplitude_final}
\begin{split}
   M_{\bm p_f} \propto -i 
   \lim_{t \rightarrow \infty} 
   \sum_{s} \left[\sqrt{\frac{2\pi i}{\partial^2 S(\bm{p}_s, \bm r_s, t_s)/\partial t_s^2}} \tilde{d}(\tilde{\bm p}_0, \bm r_0, t_s) \right.
   \\
   \left. \cross \frac{1}{\sqrt{|J_s(t)|}} e^{i S({\bm p}_s, \bm r_s, t_s)} e^{-i\frac{\pi \nu_s}{2}}
   \right],
\end{split}
\end{equation}
where $\nu_s$ is the Maslov index, to be properly introduced in the following section. The sum over $s$ is over all paths connecting the final and initial momentum and fulfilling the classical equations of motion, which follow from stationary action in time, position and momentum \cite{lai_influence_2015}
\begin{gather}
    \label{SPE1}\frac{1}{2}[\bm p_0 + \bm A(t_s)]^2 = -I_\text{p}, \\
    \label{SPE2}\dot{\bm r}_s(\tau) = \bm p_s(\tau) + \bm A(\tau), \\
    \label{SPE3}\dot{\bm p}_s(\tau) = -\nabla_{\bm r_s} V[\bm r_s(\tau)],
\end{gather}
respectively. The subscript $s$ indicates that a quantity is connected to the classical trajectory Eqs.~\eqref{SPE2}-\eqref{SPE3} given by initial momentum and time, $\bm p_0$ and $t_s$, related through Eq.~\eqref{SPE1}. In this way $s$ functions as a collective index $s=(\bm p_0, t_s)$ of the parameters uniquely identifying a given trajectory (Eq.~\eqref{SPE1} can have several solutions for a given $\bm p_0$). 

Equations \eqref{SPE1}--\eqref{SPE3} are referred to as the saddle-point equations (SPEs). The $t_s$ is usually related to the time of ionization, with the first SPE, Eq.~\eqref{SPE1}, expressing energy conservation at ionization. For a vanishing potential $\bm p_s (\tau)$ becomes constant, and Eq.~\eqref{SPE1} reduces to the SPE from the lowest order SFA \cite{lai_influence_2015, maxwell_coulomb-corrected_2017}, providing a link between the ionization time and final momentum of the electron. This is in agreement with the fact that the potential is neglected in the lowest order SFA. When considering SFA trajectories in the following, we are thus considering solutions to Eq.~\eqref{SPE1} and \eqref{SPE2} for a constant momentum value, $\bm p_0 = \bm p_f$, with $\bm p_f$ the final momentum value of the photoelectron.

$J_s$ in Eq.~\eqref{CQSFA_amplitude_final} is the Jacobian determinant in the mixed representation
\begin{equation} \label{Jacobi_det}
    J_s(\tau) = \det(\pdv{\bm p_s(\tau)}{\bm p_0}),
\end{equation}
in general evaluated at some time $\tau$ along the classical trajectory, Eqs.~\eqref{SPE2} and \eqref{SPE3}. Note that in Eq.~\eqref{CQSFA_amplitude_final} $J_s(t)$ is to be evaluated at asymptotic final time $t$, where the field is over and the electron far from the parent ion. The Jacobian matrix is also referred to as the stability matrix, since it describes the trajectory's sensitivity to the initial momentum value. At points along a trajectory where $J_s(\tau)=0$, so-called focal points, there is an insensitivity to changes in initial conditions of the trajectory in some direction. Effectively, this means that several trajectories (with different initial conditions) will pass through the same point, and we thus have a bunching of trajectories at the focal points. If the final momentum of a trajectory happens to be in a focal point, this bunching effect will be apparent in the final transition amplitude Eq.~\eqref{CQSFA_amplitude_final} and cause a breakdown of the semiclassical approximation in those points \cite{levit_hamiltonian_1977}. Surfaces in the final momentum distribution containing these focal points are known as `caustics' and are clearly visible in calculated photoelectron momentum distributions (PMDs) as a line of high transition amplitude. 

The ionization times in Eq.~\eqref{CQSFA_amplitude_final} are generally complex $t_s=t_s^{\text{Re}} + \imi t_s^{\text{Im}}$, so the integral of the action, Eq.~\eqref{action}, becomes a complex contour integral. We pick a contour composed of two parts; first we integrate from $t_s$ to the real time axis, parallel to the imaginary time axis. This part is interpreted to correspond to the tunneling of the electron out of the potential. During the integration of this first contour, the momentum is assumed to be constant. Thereafter, the action is integrated from $t_s^{\text{Re}}$ to infinity, corresponding to propagation in real time. In total, the action, Eq.~\eqref{action}, is thus split into tunneling and propagation parts \cite{Maxwell_Treating_2018}. In the tunneling part of the action, an integral over the potential appears. In practice this will be replaced by a Coulomb correction factor \cite{perelomov_ionization_1967,bisgaard_tunneling_2004,brennecke_gouys_2020}:
\begin{equation}\label{tunnel_integral_substitution}
    \exp\qty[-\int_{t_s}^{\Re(t_s)} \dd{\tau}V[\bm r_s(\tau)]]\longrightarrow \qty(
    \frac{2\kappa^2}{|\bm E(t_s)|})^{\frac{2}{\kappa}},
\end{equation}
where $\kappa=\sqrt{2I_p}$. The reasoning for this substitution comes from Perelomov-Popov-Terent’ev (PPT) theory \cite{perelomov_ionization_1967, bisgaard_tunneling_2004}, which is widely used to explain the ionization rates of atoms.  Specifically, by regularizing the integral and expanding in the path to second order, one obtains a finite integral, which resembles the expression that has been substituted. A similar result was also recently derived in a CQSFA setting in Ref.  \cite{rodriguez_forward_2023}.
In order to propagate $t\rightarrow\infty$, after the pulse is over and the electron is sufficiently far from the residual ion, we analytically propagate the equations of motion and compute the action, using the same procedure as outlined in \cite{shvetsov-shilovski_semiclassical_2016}.

%The reasoning for this is twofold: First, the potential diverges for $\bm r = 0$, and branch cuts in the potential lead to discontinuities in the final distribution \cite{pisanty_slalom_2016, Maxwell_Treating_2018}. The integral can be regularized \cite{Maxwell_Treating_2018, popruzhenko_keldysh_2014}, but defects due to the branch cuts remain.
%Therefore, we simply opt to perform the above substitution. It has also been shown, in the context of the CQSFA, that for the Coulomb potential, an approximate analytical evaluation of the exponential of the potential integral results in a expression similar to the above correction factor \cite{rodriguez_forward_2023}.

In order to calculate the ionization matrix element, Eq. \eqref{SFA_mel}, we use the asymptotic expansion of a Coulomb wave function 
\begin{equation}
\psi_0(\bm r) \xrightarrow{r\rightarrow\infty} \sum_{\ell, m} C_{\ell m} r^{Z/\kappa -1} e^{-\kappa r} Y^{m}_\ell (\theta, \phi),
\end{equation}
where $Y^m_\ell(\theta,\phi)$ is a spherical harmonic, $\kappa = \sqrt{2I_p}$ and $C_{\ell m}$ an expansion coefficient. This expansion has previously been shown to work well in the length gauge \cite{gribakin_multiphoton_1997}, also when comparing with experimental data for molecules \cite{kjeldsen_strong-field_2004}. Following the regularization procedure outlined in Ref.~\cite{gribakin_multiphoton_1997}, we find the matrix element evaluated in the SFA time saddle points, $t_s$, to be  
\begin{equation} \label{d_asymp}
\begin{split}
    d(\bm p, t_s) \propto E(t_s) (S''(t_s))^{-(1/2\kappa +1)} 
    \sum_{\ell, m} C_{\ell, m} \qty(\frac{\tilde{\bm p}(t_s)}{i\kappa})^\ell 
    \\ 
    \cross \qty(\alpha_- Y^m_{\ell-1}(\Omega_{\tilde{\bm p}}) \frac{\kappa^2}{\tilde{\bm p}(t_s)} - \alpha_+ Y^m_{\ell + 
    1}(\Omega_{\tilde{\bm p}}) \tilde{\bm p}(t_s))
\end{split}
\end{equation}
with the abbreviations
\begin{align}
    \alpha_+ &= \qty[\frac{(\ell-m+1)(\ell+m+1)}{(2\ell+1)(2\ell+3)}]^{1/2}, \\
    \alpha_- &= \qty[\frac{(\ell-m)(\ell+m)}{(2\ell-1)(2\ell+1)}]^{1/2}.
\end{align}

The constant of proportionality is neglected, since we renormalize the simulation results. For the simulations performed here, we only employ one spherical harmonic in our initial state expansion, choosing $C_{0,0}=1$ for an s-state (hydrogen and helium) and $C_{1,0}=1$ for a p-state (argon), aligned along the laser polarization direction.

Finally, we assume that the electron starts from $\bm r_s(t_s)=\bm 0$ at the start of the tunneling, in which case we can simply set the phase factor $\bm r_0 \cdot \tilde{\bm p_0}$ in Eq.~\eqref{the_d_tilde} to zero, and neglect the boundary term in the action, i.e., neglect the last term in Eq.~\eqref{action}. 

\subsection{The Maslov Phase \label{sec:theory:Maslov}}
The full CQSFA transition amplitude, Eq.~\eqref{CQSFA_amplitude_final}, contains the phase-factor $e^{-i\frac{\pi\nu_s}{2}}$ known as the Maslov phase with $\nu_s$ the Maslov index. This phase factor has not been included in previous versions of the CQSFA, but as we will show it is paramount for achieving excellent agreement with the TDSE with a semiclassical model for ATI. Recently, the Maslov phase was included in a semiclassical strong-field model \cite{brennecke_gouys_2020}, similar to the CQSFA. In this section, we elaborate on this procedure, based on Refs. \cite{levit_focal_1978, levit_hamiltonian_1977}.

First, a brief note on what the Maslov index represents. While its consequence is readily tractable from Eq.~\eqref{CQSFA_amplitude_final}, changing the phase of the transition amplitude of each trajectory, the understanding of when it arises and how to calculate it is not so simply understood. There, however, exists an analogy from optics \cite{brennecke_gouys_2020}: When focusing a Gaussian beam, it asymptotically obtains a $\pi$ phase-shift when passing through its focus, known as the Gouy phase shift. Similarly, the ionic potential may act as a focusing lens and focus a manifold of trajectories with infinitesimally small deviations in initial conditions to a single point. When and where these focal points are in space and time depend upon both the ionic potential, the properties of the laser-electron interaction potential, and consequently also on each trajectory's initial conditions. At these focal points, for which $J_s(\tau)=0$, the Maslov index may instantaneously change its value in integer steps. The Maslov phase is hence a time-dependent function, $\nu_s=\nu_s(\tau)$, and the contributions of each focal point along the trajectory must be calculated.

To calculate $J_s(\tau)$, we require the equations of motion of the stability matrices, which are in the form of a Jacobi initial-value problem. The derivation of the equations of motion and their initial conditions are detailed in Appendix \ref{app:maslov}. We find that the time-evolution of the stability matrices in the mixed representation are governed by the following coupled matrix-differential equations
\begin{gather}
   \dv{\tau} \pdv{\bm p_s(\tau)}{\bm p_0} = -\vb{H}[V(\bm r_s(\tau))]\pdv{\bm r_s(\tau)}{\bm p_0} \label{dpdp_eq} \\
   \dv{\tau}\pdv{\bm r_s(\tau)}{\bm p_0} = \pdv{\bm p_s(\tau)}{\bm p_0}, \label{drdp_eq} 
\end{gather}
where $\vb{H}[V(\bm r_s(\tau))]$ denotes the Hessian of the potential, which, as noted, must be evaluated along the classical trajectories. The initial conditions of the stability matrices read
\begin{equation} \label{JIVP_init_conds}
    \eval{\pdv{\bm p_s(\tau)}{\bm p_0}}_{\tau=\Re(t_s)}=\mathds{1}, \qquad \eval{\pdv{\bm r_s(\tau)}{\bm p_0}}_{\tau=\Re(t_s)} = \bm 0.
\end{equation}
For Eq. \eqref{potential} the Hessian reads
\begin{equation}
    \vb{H}[V(\bm r_s)] = \frac{1}{r_s}\dv{V(r_s)}{r_s}\mathds{1} - \frac{1}{r_s^3}\qty[\dv[2]{f(r_s)}{r_s} + 3\dv{V(r_s)}{r_s}] \bm r_s \otimes\bm r_s,
\end{equation}
where $\mathds{1}$ denotes the $3\times3$ identity matrix and $\otimes$ the tensor product. Propagation of Eqs.~(\ref{dpdp_eq}) and (\ref{drdp_eq}) along with the SPEs, Eqs.~(\ref{SPE1})--(\ref{SPE3}), then allows determination of $J_s(\tau)$ through Eq.~\eqref{Jacobi_det}. 

With $J_s(\tau)$ obtained for any time $\tau$ along the trajectory, we can determine the times $\tau_i$ for which $J_s(\tau_i)=0$, that is, the focal points. Since $\nu_s$ only changes in these points, its value at the final time $t$ can be written as the sum of these changes at the focal times 
\begin{equation} \label{maslov_index_tot}
    \nu_s(t) = \nu_{s,0} + \sum_i\Delta\nu_s(\tau_i),
\end{equation}
where $\nu_{s,0}$ is the initial value at the start of the trajectory, $\Delta \nu_s(\tau_i)$ the change at a given focal point (to be determined below), and the sum runs over all focal points $\tau_i \in [\Re(t_s), t]$. The initial value $\nu_{s,0}$ is only nonzero if the trajectory happens to start upon a focal point \cite{levit_focal_1978}. When $\nu_s(t)$ is to be used in the final transition amplitude, Eq.~\eqref{CQSFA_amplitude_final}, $t$ is again to be seen as an asymptotic time, $t\rightarrow\infty$, but practically large enough that the laser field is over, and the photoelectron is far from its parent ion. At the focal points, the determinant of the stability matrix will be zero, $J_s(\tau_i)=0$, and consequently $\partial \bm p_s(\tau_i) / \partial \bm p_0$ will have some number, $m$, of linearly independent eigenvectors with corresponding eigenvalues of 0, denoted $\bm d^{(k)}$, that is, $m$ solutions to the eigenvalue problem
\begin{equation}
    \frac{\partial\bm p_s(\tau_i)}{\partial\bm p_0} \bm d^{(k)} = \lambda_k \bm d^{(k)},
\end{equation}
with $\lambda_k=0$ for $k=1,\ldots ,m$. Given the $\bm d^{(k)}$ eigenvectors at these times, the change in Maslov index at a focal point $\Delta\nu_s(\tau_i)$ is given by the following formula 
\begin{equation} \label{Jacobi_IVP_delta_nu_thing}
    \Delta\nu_s(\tau_i) = m - 1 + \text{sgn}\det\bm\Pi(\tau_i), 
\end{equation}
where $\bm\Pi(\tau_i)$ is a matrix with the following components
\begin{equation} \label{Jacobi_IVP_pi_thing}
    \Pi_{nm}(\tau_i) = \bm u^{(n)}(\tau_i)\cdot \vb{H}[V(\bm r_s(\tau_i))]\bm u^{(m)}(\tau_i),
\end{equation}
where we have defined 
\begin{equation} \label{Jacobi_IVP_s_thing}
    \bm u^{(k)}(\tau_i) = \pdv{\bm r_s(\tau_i)}{\bm p_0} \bm d^{(k)}.
\end{equation}
From our experience, setting $\nu_{s,0}=0$ for all cases will not notably change the result due to the occurrence of focal points at the start of the propagation being extremely rare.

\subsection{The inversion problem \label{sec:theory:Inversion}}\label{sec:inversion}
\begin{figure}
    \centering
    \includegraphics[width=\linewidth]{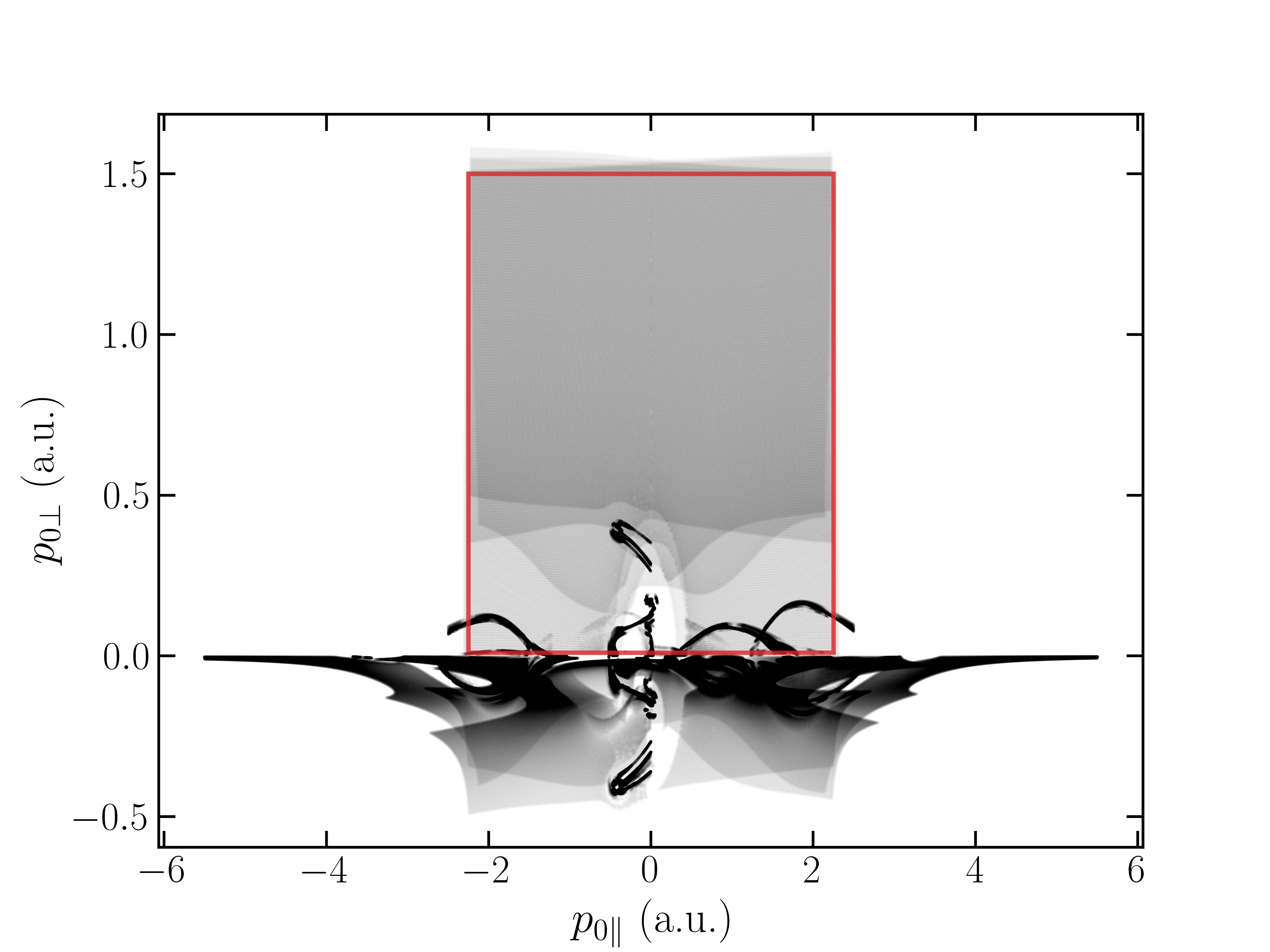}
    \caption{Initial momentum values sampled for a simulation of helium ionized by a two cycle $\sin^2$ laser pulse with intensity $4\cross 10^{14}$ W/cm$^2$ and wavelength of 800 nm. The darker color indicates high density of sampled points. The red box marks the region of final momentum values investigated in the simulation, i.e., all the shown initial conditions lead to a final momentum value within this box. Note the difference in scale between the $p_{0\parallel}$ and $p_{0\perp}$ axis. Figs.~\ref{fig:validation_PMDs}(b) and (e) shows the corresponding PMD.}
    \label{fig:init_conds}
\end{figure}

In order to calculate the transition amplitude, Eq.~\eqref{CQSFA_amplitude_final}, we sum over all saddle-point solutions fulfilling Eq.~\eqref{SPE1}--\eqref{SPE3}. Given that this transition amplitude is formulated via a mixed representation path integral, this means we have an initial condition $\bm r_s(t_s)=0$ and a boundary condition $\bm{p}_s(\tau\rightarrow\infty)\rightarrow\bm{p}_f$. Thus, to find solutions, one must find all the initial momenta $\bm{p}_0$ that lead to a particular final momentum $\bm{p}_f$. This is known as the inversion problem. Due to the difficulty of finding all of these solutions, many models take an alternative approach, where many starting trajectories are forward propagated and the final momentum binned, so that trajectories that end in the same final bin are summed coherently. To resolve interference, this requires very small bins and typically in excess of a billion trajectories. Additionally, it has also been shown in Ref.~\cite{shvetsov-shilovski_semiclassical_2021} that these `forward' approaches do not yield the correct sampling weight in terms of the Jacobian $J_s$, leading to $1/|J_s|$ instead of the correct $1/\sqrt{|J_s|}$ computed by inverse approaches (see Eq.~\eqref{CQSFA_amplitude_final}).

In earlier iterations of the CQSFA \cite{maxwell_coulomb-corrected_2017,maxwell_coulomb-free_2018,maxwell_analytic_2018}, not all solutions to the SPEs were included. This method assumed that there were only 4 solution per laser cycle and used an iterative procedure to find them. Later, additional solutions were identified \cite{maxwell_coulomb-free_2018} but were not incorporated into the CQSFA. The approach taken here, is to use a Monte-Carlo-style sampling method, to find all the initial momenta that lead to any particular final momentum, see Fig.~\ref{fig:init_conds}. This works by generating a \textit{partially} random guess for an initial momentum $\bm{p}_{0,g}$, which (propagating via the SPEs) leads to a particular final momentum $\bm{p}_{f,g}(\bm{p}_{0,g})$. Then, in analogy to the `forward' method, we bin this solution, associating it with the closest final momentum grid point $\bm{p}_{f,j}$ to $\bm{p}_{f,g}(\bm{p}_{0,g})$, where $j$ denotes the index of the grid. 
However, now this method differs from the `forward' method, as we change the initial momentum to make the resulting final momentum approach that of the grid point.
More precisely, the initial guess momentum $\bm{p}_{0,g}$ is used as a starting point to find a slightly different initial momentum $\bm{p}_{0,g}'$ so that the final guess momentum and the grid point momentum are essentially equal, by solving $|\bm{p}_{f,g}(\bm{p}_{0,g}')-\bm{p}_{f,j}|<\epsilon$, where $\epsilon$ is taken to be a very small value of the order $\epsilon\sim 10^{-12}$.
When sampling several initial guesses, care must be taken that any new initial guesses $\bm{p}_{0,g}$ do not lead to a solution that has already been binned.

If enough initial momenta are provided by a random number generator over a large enough range of $\bm{p}_0$, all unique solutions can be found in this way. However, this algorithm can be very inefficient if solutions are clustered very densely in small areas of $\bm{p}_0$ momentum space, which is the case for rescattered trajectories. An example of the set of all initial momentum obtained for a Helium simulation is given in Fig.~\ref{fig:init_conds}. The rescattered trajectories have a negative initial momentum coordinate perpendicular to the laser field polarization $p_{0\perp}<0$, which form densely clustered points that are more difficult to sample. To address this, we implemented an algorithm that can detect narrow regions of densely clustered solutions, and then sampling can be more concentrated in these regions. These regions can be detected after performing an initial sampling over the full region depicted in \figref{fig:init_conds}. Then a grid is placed over the region $p_{0\perp}<0$, which is coarse enough so that if a dense cluster of points intersects a box in the grid, the initial sampling will have found at least one solution in this box. The clusters can then be intensively searched by restricting guesses only to boxes in the grid with at least one solution already present.

A similar approach was recently implemented in Ref.~\cite{rodriguez_forward_2023}, however, in this case the sampling was done with a Gaussian distribution. This is efficient for the direct trajectories that do not undergo rescattering, but less so for the rescattered trajectories. Ref.~\cite{rodriguez_forward_2023} also did not consider a pulse envelope or the Maslov phases that we consider here.

\subsection{Orbit types \label{sec:theory:orbits}}

\begin{figure}
    \centering
    \includegraphics[width=\linewidth]{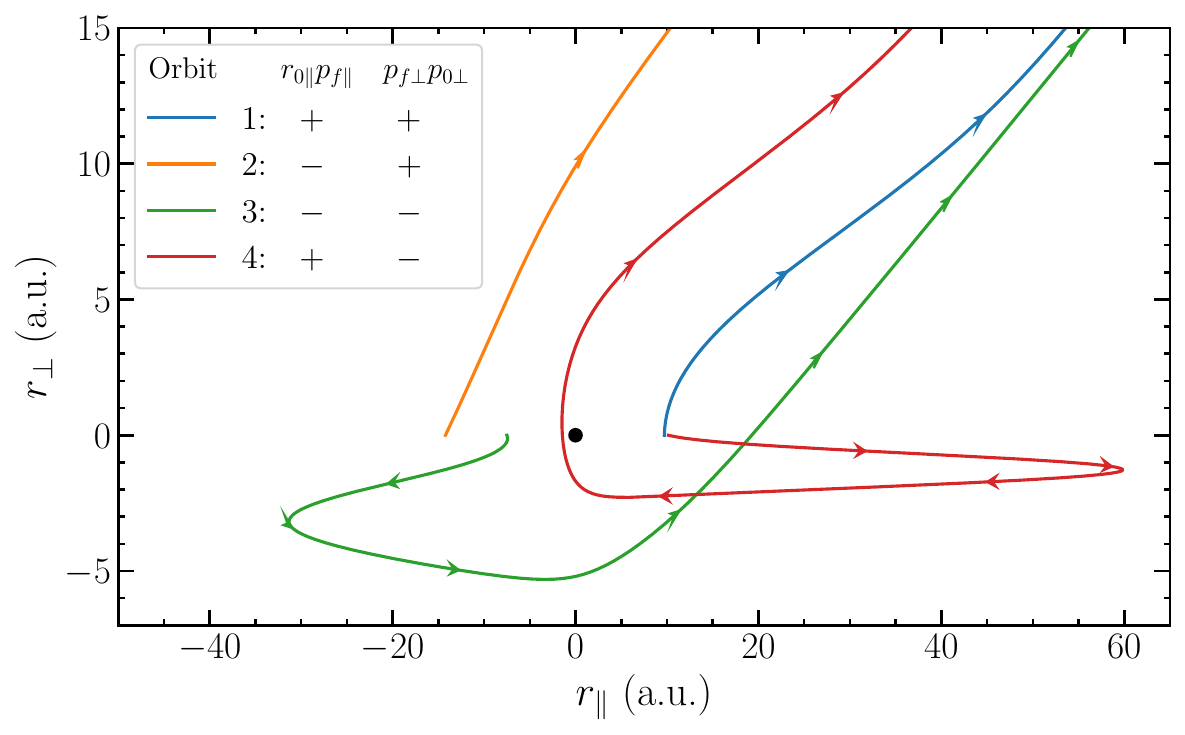}
        \caption{An example of the four different orbit types in the CQSFA for an asymptotic final momentum $\bm p_f = (0.6, 0.25)$ a.u., taken from the helium simulation of Fig.~\ref{fig:init_conds}. The arrowheads indicate the direction of travel of the photoelectrons and are separated by 0.1 laser pulse periods in time. The legend indicates the definition of each orbit type in terms of the sign of the product between initial position and final parallel momentum, and final and initial perpendicular momentum. The trajectories shown here are the ones usually captured in previous CQSFA models.}
    \label{fig:orbits}
\end{figure}

In previous versions of the CQSFA the division of the quantum orbits into 4 types \cite{yan_low-energy_2010} was explicitly used in the simulations \cite{lai_influence_2015, maxwell_coulomb-free_2018}. For a laser linearly polarized along the $z$ direction, the classification compares the tunnel exit 
\begin{equation}
    z_{0} = \int_{t_s}^{\Re(t_s)} A(\tau) d\tau,
\end{equation}
and the final momentum along the polarization direction, $p_{f\parallel}$, to the final and initial perpendicular momentum $p_{f\perp}$ and $p_{0\perp}$, through the signs of the products $r_{0\parallel} p_{f\parallel}$ and $p_{f\perp} p_{0\perp}$, as summarized in the legend of Fig.~\ref{fig:orbits}. We briefly describe the `standard' trajectories of each orbit type, also shown on the same figure. Orbit type 1 ($r_{0\parallel} p_{f\parallel} > 0$, $p_{f\perp} p_{0\perp}>0$) corresponds to direct trajectories, not revisiting the ion core. Type 2 ($r_{0\parallel} p_{f\parallel} < 0$, $p_{f\perp} p_{0\perp}>0$) describes trajectories driven past the core by the field, but without much interaction. Type 3 ($r_{0\parallel} p_{f\parallel} < 0$, $p_{f\perp} p_{0\perp}<0$) corresponds to trajectories performing a field-driven forward scattering of the ion core. Last, type 4 ($r_{0\parallel} p_{f\parallel} > 0$, $p_{f\perp} p_{0\perp}<0$) describes trajectories performing a field-driven backwards scattering of the core. Orbit types 1 and 2 only interact weakly with the ion potential, and can thus be related to the trajectories found in the lowest order SFA (see the text following Eq.~\eqref{SPE3}). On the other hand, orbit type 3 and 4 require the presence of the Coulomb potential. Since the semiclassical trajectories here are restricted to a plane we use a 2D definition of the orbit types, but a simple generalization to 3D trajectories is possible \cite{maxwell_relativistic_2023}.

We note that other trajectories than these `standard' versions exist, such as multi-pass trajectories, having multiple close encounters with the ion core, and directly re-colliding trajectories, where the ionization happens at low fields and the scattering is primarily driven by Coulomb forces. While these were neglected in previous CQSFA versions \cite{maxwell_coulomb-free_2018}, the current version of the CQSFA will include all trajectory dynamics, assuming sufficient sampling.

\section{Results and discussion \label{sec:Results}}
\begin{figure*}
    \centering
    \includegraphics[width=\linewidth]{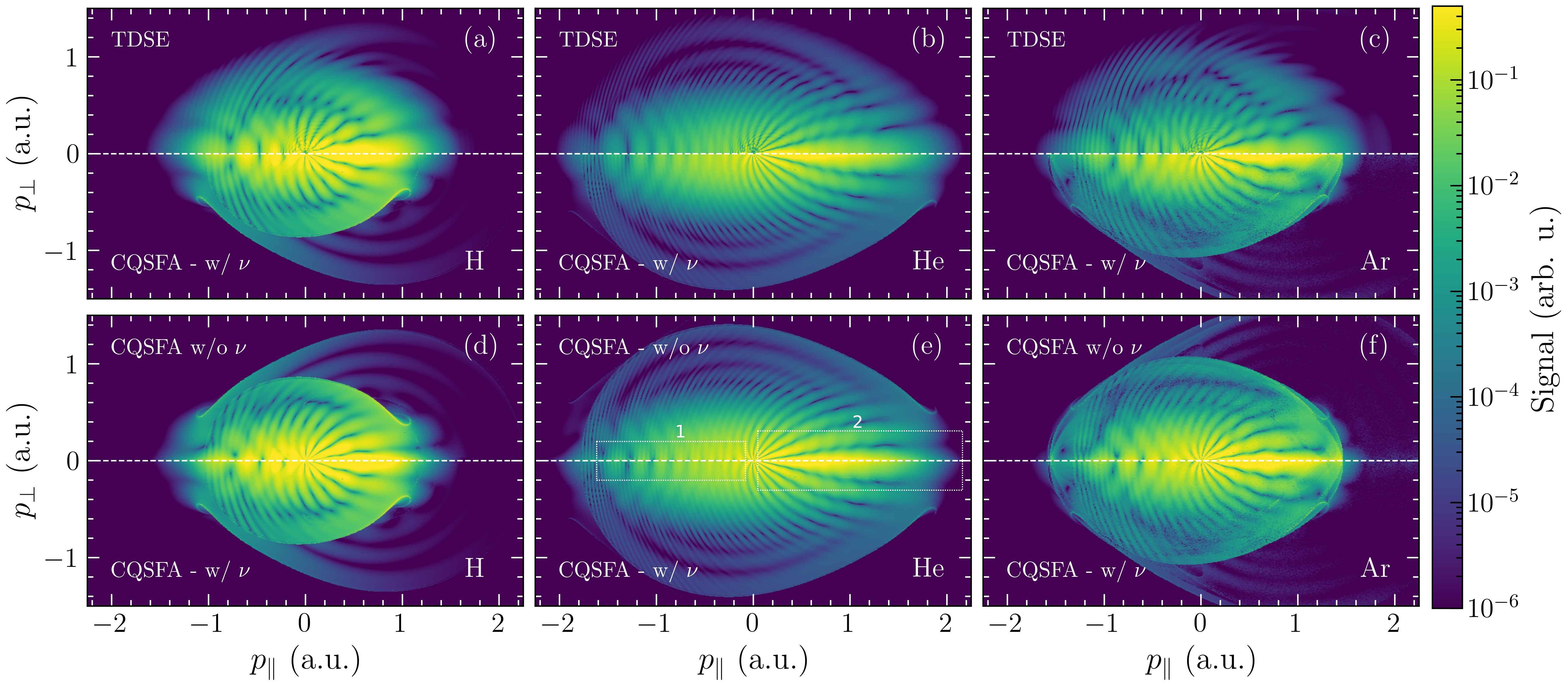}
    \caption{Photoelectron momentum distributions for H, He and Ar subject to a two-cycle sin$^2$-pulse with a carrier-envelope phase set to $\varphi=\pi/2$ [see Eq.~\eqref{A_field}]. The wavelength is set to 800 nm, corresponding to $\omega=0.057$, for all calculations. Due to the cylindrical symmetry, $[H(t),L_\parallel]=0$, only half of the PMD is required. The horizontal dashed white lines in each subplot signify the boundary between the upper and lower parts. In the first row, we compare TDSE results (upper half) for H (a), He (b) and Ar (c) to the improved CQSFA, including the Maslov phase (lower half). In the second row, we then compare the CQSFA calculations with (without) the Maslov phase in the bottom half (top half) for H, He and Ar in (d), (e) and (f), respectively. The intensity of the laser is for each target given by $1.3\times10^{14}\,\text{W}/\text{cm}^2$ for H, $4\times10^{14}\,\text{W}/\text{cm}^2$ for He and $2\times10^{14}\,\text{W}/\text{cm}^2$ for Ar. These parameters ensure that we are within the tunneling regime and the Keldysh parameter $\gamma_\text{K}$ satisfying $\gamma_\text{K}<1$ for all targets. The single colorbar is representative for all PMDs.}
    \label{fig:validation_PMDs}
\end{figure*}

In this section, we present and discuss results obtained with the model and compare to calculations performed using the freely available time-dependent Schrödinger equation (TDSE) solver, \Qprop\ \cite{Qprop_2019}. \Qprop\ works within the electric dipole and SAE approximation, and can calculate photoelectron spectra using the i-SURFV method. The radial step size and time step used in simulations were adjusted to ensure the correct ground state energy in imaginary time propagation and convergence of the PMDs, down to $\Delta r=0.01$ a.u.\ and $\Delta t=0.0025$ a.u.\ for argon. The size of the angular grid was chosen as $r_{max}=100$ a.u.\ with the t-SURFF boundary at $300$ a.u.\ Angular momenta up to $\ell =100$ were considered for argon and hydrogen, and up to $\ell =150$ for helium. 

For all calculations, both the TDSE and CQSFA, the coefficients for the effective SAE potentials for Eq. \eqref{f_function} are taken from Ref.~\cite{tong_empirical_2005}. Throughout we take the vector potential to be a linear sin$^2$-field of the form
\begin{equation} \label{A_field}
    \bm A(t) = 2\sqrt{U_\text{p}}\sin^2\qty(\frac{\omega t}{2 N_c})\cos(\omega t + \varphi) \hat{z},
\end{equation}
where $\hat{z}$ is a unit vector in the $z$ polarization direction, $U_\text{p}$ is the ponderomotive potential, $\omega$ is the carrier frequency of the field, $N_c$ is the number of cycles and $\varphi$ is the carrier-envelope phase (CEP). All the results in the following sections will be for a 800 nm ($\omega=0.057$ a.u.) $N_c=2$ or $4$ cycle pulse, with $\varphi=\pi/2$. The intensity was varied for each target, using $1.3 \cross 10^{14}$ W/cm$^2$ ($U_\text{p}=0.29$ a.u., $\gamma_\text{K}=0.94$) for hydrogen, $4 \cross 10^{14}$ W/cm$^2$ ($U_\text{p}=0.88$ a.u., $\gamma_\text{K}=0.72$) for helium and $2 \cross 10^{14}$ W/cm$^2$ ($U_\text{p}=0.44$ a.u., $\gamma_\text{K}=0.81$) for argon, to ensure low Keldysh parameters, $\gamma_\text{K}=\sqrt{I_\text{p}/2U_\text{p}}$, without having over the barrier dynamics. 

The systems here all have cylindrical symmetry. Therefore, it is sufficient to consider a 2D slice of the results, so we consider only results in the $zy$-plane for $x=0$. Furthermore, the symmetry causes all the semiclassical trajectories to be confined within a plane, meaning sampling over initial momenta also only has to be performed in the $zy$-plane. We note, however, that the calculations are performed with a 3D model which has implications for determining the correct value of $J_s(\tau)$, Eq.~\eqref{Jacobi_det}, \cite{brennecke_gouys_2020} and also allows for calculations for systems without cylindrical symmetry. In the following sections we will refer to the $y$ and $z$ coordinates as $r_\perp$ and $r_\parallel$, the coordinate perpendicular and parallel to the laser polarization respectively. Similarly, we denote the parallel ($p_z$) and perpendicular ($p_y$) momentum components $p_\parallel$ and $p_\perp$. Unless otherwise stated, the momenta used in the following will all refer to the final momenta of the semiclassical trajectories. 

\subsection{Validation of the model}\label{sec:validation}
In Fig.~\ref{fig:validation_PMDs}, we compare CQSFA and TDSE simulations performed for hydrogen, helium, and argon. The simulation parameters for helium are the same as used in Ref.~\cite{brennecke_gouys_2020}, in order to enable direct comparison. For all targets, we see a very good agreement between the TDSE and the CQSFA with the Maslov phase included, Fig. \ref{fig:validation_PMDs}(a)--(c), in the lower energy region before the presence of caustics. As mentioned in the section following Eq.~\eqref{Jacobi_det}, the caustics are surfaces containing focal points for the classical trajectories at the final asymptotic time. Because of the breakdown of the semiclassical model in these points, the caustics are generally visible in the PMDs as sharp lines of discontinuity in the signal, e.g. in Fig.~\ref{fig:validation_PMDs}(a) at $\bm p \sim (0, -0.8)$ a.u. 

When we include the Maslov phase into the CQSFA, it is evident that especially the location of the interference fringes along $p_\parallel$ at $p_\perp\sim0$ are very well-reproduced. The fact that the Maslov phase is indeed necessary to obtain this agreement is illustrated in Fig.~\ref{fig:validation_PMDs}(d)--(f). Here, the CQSFA with and without the Maslov phase is shown on each half, and a shift of interference fringes is clearly seen along the $p_\perp=0$ axis, see panel (e) box 1 for He. 

\begin{figure}
    \centering
    \includegraphics[width=\linewidth]{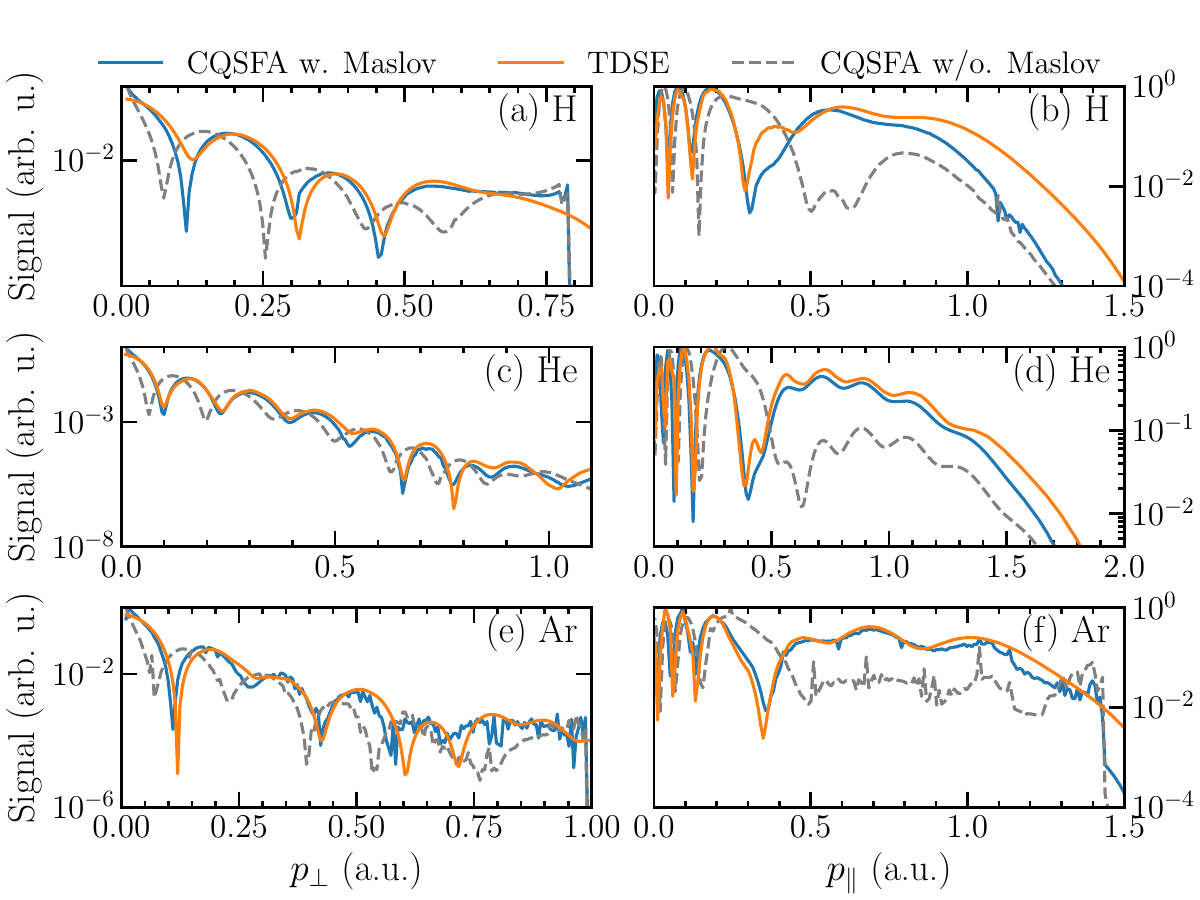}
    \caption{Transition amplitude signal for the TDSE simulations (orange lines) and the CQSFA with Maslov phase (blue line) and without (gray striped line), through different slices of the PMDs in Fig. \ref{fig:validation_PMDs}. Left column [(a), (c), (e)] shows the signal for $p_\parallel=0.4$ (a.u.) for hydrogen, helium and argon respectively. The right column [(b), (d), (f)] shows the signal for $p_\perp=0.1$ (a.u.) for the same order of elements. The same laser parameters as in Fig.~\ref{fig:validation_PMDs} are employed. All signals are normalized according to their respective maximum values in the plotted region.}
    \label{fig:slice_plots}
\end{figure}

Another important structure affected by the inclusion of the Maslov phase is the fringes at $p_\parallel > 0$ almost parallel to the $p_\parallel$ axis, commonly known as the `spider-legs' \cite{faria_it_2020}. This interference pattern is found for a wide range of targets, and is visible in all the PMDs in Fig.~\ref{fig:validation_PMDs}, most clearly for He. As seen on panels (d)--(f) the Maslov phase results in a `lift' of the central fringe of the spider (panel (e) box 2 for He), increasing its width to a size in much better agreement with the TDSE results. This effect is further illustrated in Figs.~\ref{fig:slice_plots}(b), (d), and (f), showing the signal along a slice of the PMDs of Fig.~\ref{fig:validation_PMDs} for $p_\perp=0.1$ a.u. While the CQSFA results both seem to agree with the TDSE simulations for low $p_\parallel$, the inclusion of the Maslov phase clearly creates a difference for $p_\parallel > 0.5$ a.u., increasing agreement with the TDSE results significantly. In Figs.~\ref{fig:slice_plots}(a), (c), and (e), a slice of the signal along $p_\parallel=0.4$ a.u.\ also illustrates how the location of the other spider legs are affected by the Maslov phase. For all targets the agreement is better with the inclusion of the Maslov phase, perhaps most clearly for hydrogen, where quite large shifts in the interference minimum can be seen (compare e.g. the gray dashed line at $p_\perp=0.25$ a.u.\ to the blue at $p_\perp=0.30$ a.u.\ in Fig.~\ref{fig:slice_plots}(a)). This illustrates the necessity to take the Maslov phase into account, if one wants to perform analysis based on the spider structure. 

Still, the CQSFA and TDSE results do not agree completely. One particular region where the results seem to disagree, is for high values along the $p_\parallel$ axis, where the CQSFA results drop off too fast. This is clearly illustrated for both hydrogen and helium in panels (b) and (d) in Fig.~\ref{fig:slice_plots}, and can also be seen in the PMDs of Fig.~\ref{fig:validation_PMDs}. The disagreement seems to be worst for hydrogen. The reason for this could be the fact that the hydrogen simulation is performed for the lowest intensity of all the simulations, and in connection also the simulation with the highest Keldysh parameter ($\gamma_\text{K} = 0.94$). To test the range of validity of the CQSFA, we performed several simulations with values of Keldysh parameters in the range $[0.7, 1.1]$. In general, testing suggests that the CQSFA becomes better for longer wavelengths \cite{maxwell_relativistic_2023}, higher intensities and lower Keldysh parameters, a trend in conformity with the well performing helium simulation which has the highest intensity and lowest Keldysh parameter ($\gamma_\text{K}=0.72$). The CQSFA also performs well for intensities corresponding to over-the-barrier dynamics. 

Another source of discrepancy between the CQSFA and TDSE results is, as mentioned, the caustics. The effect of those are worse for the hydrogen and argon simulations, as seen in panel (a) and (c) of Fig. \ref{fig:validation_PMDs}, where they cut off most of the structure at high energies, an effect also visible as a sharp drop-off on panels (a) and (f) of Fig.~\ref{fig:slice_plots}. Since the caustics represent classical turning points, their position can be pushed outwards from the center by increasing, for example, the intensity or the number of cycles of the pulse. 

The results also show that the argon simulation is less converged than the hydrogen and helium simulations, see Figs.~\ref{fig:slice_plots}(c) and (f), where the blue and gray lines showing the Ar CQSFA results are clearly less smooth than the corresponding curves for He and H, Figs.~\ref{fig:slice_plots}(a)--(d). The reason for this is that the effective potential used for argon gives rise to many additional (and seemingly irrelevant) solutions to the SPEs, meaning significantly  more sampling of orbits is required to reach convergence. It is therefore worth noting, that the inclusion of $f(r)$ in Eq.~\eqref{potential} is not critical for the current simulations. Just using the plain Coulomb potential (but the correct $I_p$ for the targets) yields results almost identical to those in Fig.~\ref{fig:validation_PMDs}, with some small difference in the signal value for the high energy rings, while convergence is reached with less sampling. It could be, however, that the effective potential plays a larger role for other systems or field parameters, such as fields with more cycles where higher energies are reached.

Overall, the large degree of agreement between the CQSFA results and the TDSE simulations shown in Fig. \ref{fig:validation_PMDs} illustrates the accuracy and validity of the semiclassical model.

\subsection{Trajectory dynamics} \label{sec:orbit_types}

\begin{figure*}
    \centering
    \includegraphics[width=0.49\linewidth]{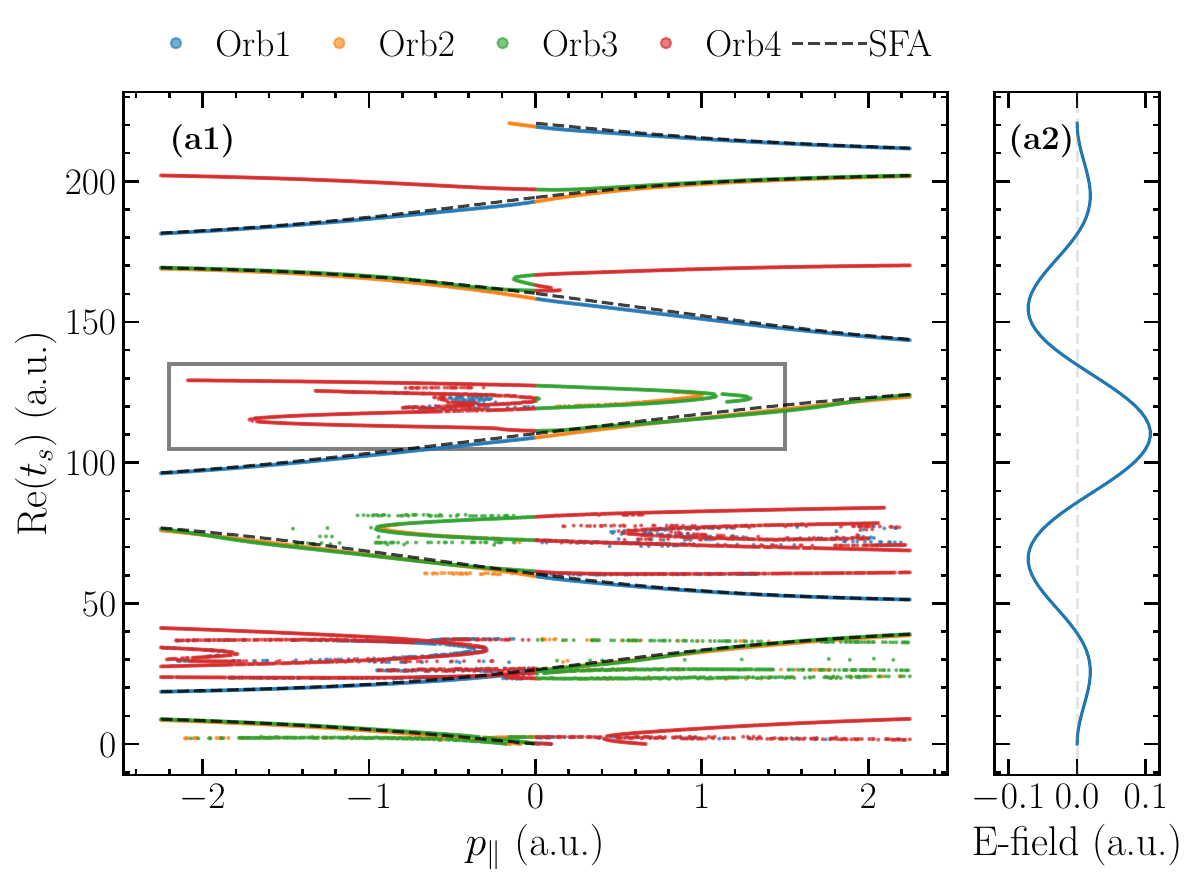}
    \includegraphics[width=0.45\linewidth]{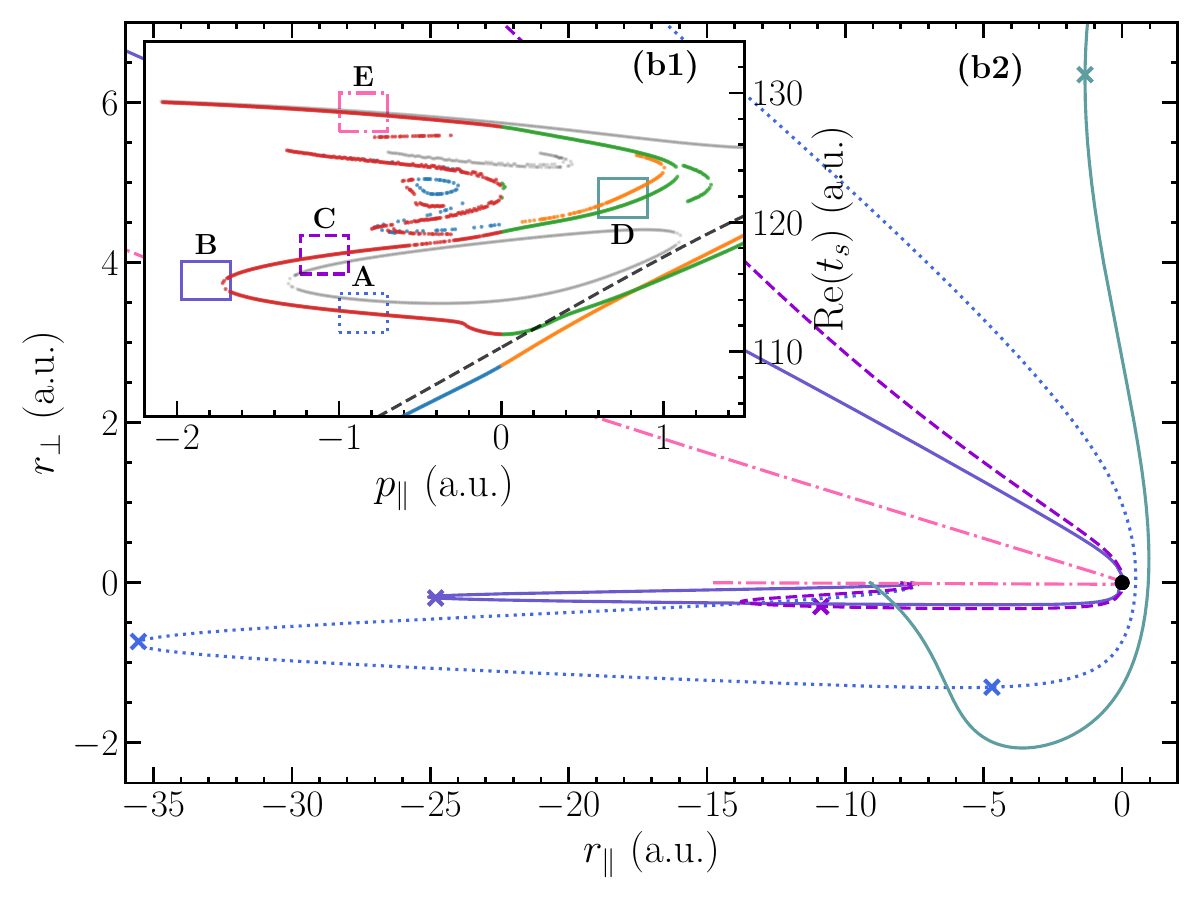}
    \caption{(a1): Real time of the temporal saddle points as a function of final parallel momentum $p_\parallel$ for the orbits used in the helium simulation of Fig.~\ref{fig:validation_PMDs} (2-cycle $\sin^2$ pulse with intensity of $4\cross 10^{14}$ W/cm$^2$ and wavelength of 800 nm) at $p_\perp=0.25$ a.u. The different orbit types (1--4) are plotted in different colors, and the dashed line indicates the ionization times obtained using standard SFA for the same final momentum values. (a2): Electric field shown as a function of the real part of the ionization time. (b1): Zoom-in of orbits marked with a gray box in panel (a1). The light gray orbits in this panel are orbit 3 and 4 for $p_\perp=1$ a.u.\ and are not related to the other panels. (b2): Real space trajectories of one orbit from each box in (b1). The trajectory color and line style match the corresponding box in the insert. Note that from box D an orbit 3 trajectory is plotted. The crosses mark points along the trajectories with electric field maxima.}
    \label{fig:saddle_point_plot}
\end{figure*}

The high level of accuracy of the CQSFA adds a new level of credibility to orbit-based analyses. Given how well the semiclassical model reproduces the TDSE results, it seems likely that the sampled trajectories and their properties could shed light onto key aspects of the underlying physics. Studying the different types of trajectories found in the new model, along with their impact on the PMDs, could thus lead to an improved understanding of the ATI process. 

To investigate the different types of trajectories and their effects on the transition amplitude, we plot in Fig.~\ref{fig:saddle_point_plot}(a1) the ionization time as a function of final parallel momentum for orbits sampled in the helium simulation shown in Fig.~\ref{fig:validation_PMDs}(b), at $p_\perp=0.25$ a.u.\ (the CQSFA result in Fig.~\ref{fig:validation_PMDs}(b) is shown for $p_\perp<0$, but because of the cylindrical symmetry the result for $p_\perp>0$ is identical). Each vertical slice in the plot shows all saddle point times for a particular final momentum point in the PMD.
%, with signal from each orbit in the slice interfering to create the final transition amplitude. 
The corresponding SFA solutions (see the text following Eq.~\eqref{SPE3} for definition) are shown in dashed black lines. Clearly, the majority of orbit 1 (blue) and 2 (orange) solutions lie along the SFA solutions, indicating their close resemblance to the SFA trajectories. Between each pair of SFA `lines' one sees additional structure composed of orbit type 3 (green) and 4 (red). These structures contain the rescattering trajectories, which will be studied in detail.% in the following. 

%TRY TO FIX THIS 
%Trajectories with different dynamics are represented by different structures in the plot, e.g. the orbits 1 and 2 along the SFA lines are (unsurprisingly) similar to the SFA type orbits, while the `scattered' structures represent multi-pass trajectories with high sensitivity to initial conditions, thus being hard to sample, but also highly suppressed by the stability factor.

The rescattering structure with the most impact on the transition amplitude is the framed structure in Fig.~\ref{fig:saddle_point_plot}(a1), containing orbits with ionization times near the highest peak of the electric field. A zoom-in of this structure is seen in panel (b1) and a collection of corresponding trajectories displayed in panel (b2), to show the evolution in trajectory dynamics that the structure represents. The lowest orbit 4 line, labeled by box A, contains field-driven scattering trajectories; the trajectory has a turning point at the maximum of the electric field, then it returns to the ionic core and scatters after spending around one laser cycle in the field. These orbits correspond to the `standard' orbit 4 in Fig.~\ref{fig:orbits}, and were used in previous CQSFA models (see, e.g., Fig.~16 in Ref.~\cite{faria_it_2020}). Box B marks a turning point in the orbit 4 structure. Here the field maximum occurs after the trajectories turning points and the return to the ionic core now happens within the same cycle as ionization. In this way, the orbits from boxes A and B are similar to the `long' and `short' orbit from the rescattered SFA \cite{figueirademorissonfaria_highorder_2002} and HHG \cite{lewenstein_phase_1995}. As we move to ionization times further from the electric field maximum, these short orbits behave less like the SFA orbits, as the Coulomb interactions become more dominant, making the trajectory turning point happen sooner compared to the field maximum, see the trajectory from box C. At box D the scattering happens before the next field maximum, while at box E we have rescattering before the laser amplitude changes sign. Interestingly, and quite unlike normal SFA orbits, these orbits are ionized at times approaching 0 electric field. For the same reason, however, their contribution to the transition amplitude is small. It is possible that these trajectories play a more prominent role for more complex targets or laser fields. In such cases, they could be an interesting probe of the system potential with which they interact strongly.

It is worth noting that other trajectory types than the ones just discussed are present in Figs.~\ref{fig:saddle_point_plot}(a1) and (b1). For example, the orbit 1 and 4 structure found between box C and E on Fig.~\ref{fig:saddle_point_plot}(b1) represents multipass trajectories, rescattering with the parent ion several times. Even `chaotic' trajectories, where the electron moves in a complicated and unpredictable manner, can be found in the less connected parts of Fig.~\ref{fig:saddle_point_plot}(a1), primarily for lower times ($\Re(t_s) \in [0, 90]$ a.u.), since these trajectories needs time interacting with the field to develop. Due to their extreme sensitivity to initial conditions, however, these `chaotic' trajectories should have a large value of $J_s$, Eq.~\eqref{Jacobi_det} and in turn a small contribution to the final transition amplitude. 

We also note that the orbit structure shown in Fig.~\ref{fig:saddle_point_plot} can become more complicated for other potentials. One example is the argon potential used to obtain the results in Fig.~\ref{fig:validation_PMDs}(c) and (f), where a lot of additional solutions become possible, complicating the sampling as previously mentioned. This however also shows the possibility for additional types of trajectory dynamics to be involved in the description of more complicated targets. 

\subsection{Orbit-resolved PMDs} \label{sec:orbit_res_pmd}
In this section, we present and analyze PMDs made from only one or two orbit types. This is the first time that such PMDs have been produced with the CQSFA for few cycle pulses. As will be seen, the structures in the orbit-resolved PMDs can be directly related to Fig.~\ref{fig:saddle_point_plot} panel (a1) and especially (b1). This comparison is aided by the fact that the turning points in the orbit structure (e.g. box B in panel (b1)) correspond to a point on a caustic in the PMDs. This is the case since $\Re(t_s)$ is parameterized by the initial momentum $\bm{p}_0$, see Eq.~\eqref{SPE1}. Thus, a turning point in $\Re(t_s)$ can be expressed as 
\begin{align}
\frac{\partial \Re(t_s)}{\partial \bm{p}_f}
&= \frac{\partial \bm{p}_0}{\partial \bm{p}_f} \frac{\partial \Re(t_s)}{\partial \bm{p}_0}
%\notag\\
%&=
%\frac{\partial \bm{p}_0}{\partial \bm{p}_f}
%\frac{-(\bm{p}_0+\bm{A}(t_s))\cdot \hat{\bm{p}}_0}{\partial^2 %S(\tilde{\bm p}_s, \bm r_s, t_s)/\partial t_s^2}
\end{align}
Hence, a focal point, $J_s(t)=0$, and a caustic, will lead to a divergent turning point in $\Re(t_s)$, as can be observed in Fig.~\ref{fig:saddle_point_plot}(b1) box B. 
%It can also be determined that $\frac{\partial t_s}{\partial \bm{p}_0}=(\bm{p}_0+\bm{A}(t_s))\cdot \hat{\bm{p}}_0/S''(t_s)$.
%thus have a continues change in initial momenta. The turning points in the orbit structure therefore represents trajectories ending in the same final momenta, regardless of small variations in the initial momentum, i.e. they mark the presence of a focal point. 

\begin{figure*}
    \centering
    \includegraphics[width=\linewidth]{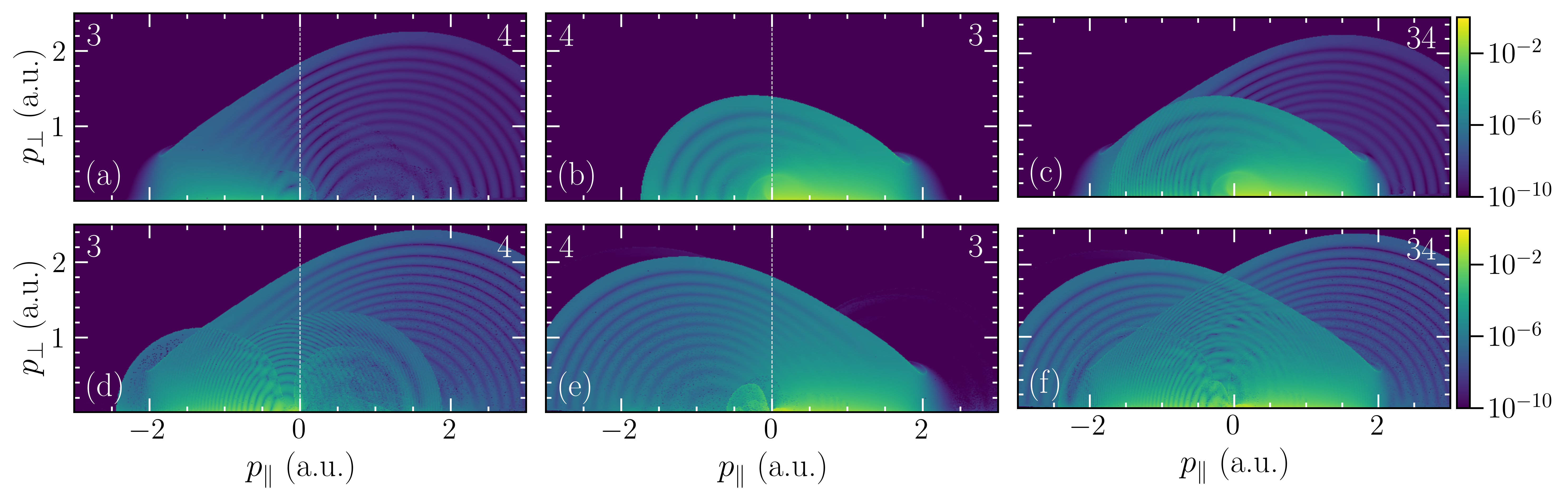}
    \caption{Orbit resolved PMDs for He of orbit type 3 and 4. Panel [(a), (b), (c)] shows the results for a two-cycle pulse, parameters identical to that of Fig. \ref{fig:validation_PMDs}. Panel [(d), (e), (f)] shows results for a 4-cycle pulse, with all other parameters identical. Panels [(a), (b), (d), (e)] shows the PMDs for the single orbits 3 and 4, with each half of a plot, separated by a dashed line, corresponding to the orbit type indicated by the number in the top corner. Panels [(c), (f)] shows the PMDs for the orbit 3 and 4 solutions combined, i.e., the orbit 3 and 4 signal is summed coherently before taking the norm square.}
    \label{fig:He_single_orbs}
\end{figure*}

In Fig.~\ref{fig:He_single_orbs}, the PMDs for orbit types 3 and 4 for a 2-cycle and 4-cycle pulse is shown. As we already saw in Fig.~\ref{fig:saddle_point_plot}, the somewhat arbitrary definition of the orbit types, mean that orbit types 4 and 3 changes into the other at $p_\parallel=0$. In order to get a continuous PMD, we stitch together the contributions of the two orbit types on each half of the momentum distribution about $p_\parallel$, as indicated in the panels. Some of the features in Figs.~\ref{fig:He_single_orbs}(a), (b) can be directly related to the distribution of $\text{Re}(t_s)$ and final parallel momentum $p_\parallel$, Fig.~\ref{fig:saddle_point_plot}(a1), with the high signal in Fig.~\ref{fig:He_single_orbs}(b) originating from the orbits in the central `branch', $\Re(t_s)\in[90, 140]$ a.u.\ as shown in Fig.~\ref{fig:saddle_point_plot}(b1), and the lower signal in (a) originating from the branch below, $\Re(t_s) \in [50,90]$ a.u. These are the only two orbit branches expected to correspond to a strong signal, since only the photoelectrons that correspond to these solutions enter the continuum at high field values, while also having adequate time to subsequently propagate in the electric field. 

We see that the momentum value where the caustic appears on the left side of Fig.~\ref{fig:He_single_orbs}(b) ($p_\parallel \sim -1.75$ a.u.), matches very well with the orbit 4 turning point in Fig.~\ref{fig:saddle_point_plot}(b1) box B, indicating that the circular interference pattern within this caustic originates from the interference of the long and short re-scattered orbit 4 (Box A and C), which is well known. The orbit 3 turning point in Fig.~\ref{fig:He_single_orbs}(b1), close to Box D, on the other hand, does not result in a visible caustic in the orbit-resolved PMD in Fig.~\ref{fig:He_single_orbs}(b) on the right-hand side (at $p_\perp = 0.25$ a.u.). This indicates that the upper part of this main structure (containing box E), associated with Coulomb-driven recolliding orbits 3 and 4 solutions, correspond to a very small transition amplitude. This situation changes at $p_\perp\sim0.6$ a.u., where a caustic appears on the orbit 3 side of Fig.~\ref{fig:He_single_orbs}(b) ($p_\parallel \sim 1.75$ a.u.). This is due to a slight change in the structure from Fig.~\ref{fig:saddle_point_plot}(b1), where at $p_\perp > 0.6$ a.u.\ the orbit 3 line near box D detaches from the orbit 3 solution above and instead connects to the orbit 3 solution below that lies along the SFA solution, forming a new loop structure together with the orbit 4 solutions in Box A, B, and C. This is shown by the light gray orbits in the same panel, indicating the orbit 3 and 4 structure for $p_\perp=1$ a.u. The formation of this ring structure increases the importance of the orbit 3 solutions around the right turning point, both leading to the caustic at $\bm p \sim (1.8, 0.6)$ a.u.\ and the appearance of interference fringes above $p_\perp \sim 0.6$ a.u.\ on the orbit 3 side of Fig.~\ref{fig:He_single_orbs}(b). 

The difference in transition probability in panels (a) and (b) of Fig. \ref{fig:He_single_orbs} is clearly an effect of the two-cycle pulse used. Indeed, in PMDs for a 4-cycle pulse, as can be seen on panels (d) and (e) of the same figure, the value of transition amplitude for a given orbit type is about the same for both $p_\parallel>0$ and $p_\parallel<0$. The effect of 2 extra field cycles is largest for orbit 4, where the high-energy interference rings are now present for both positive and negative $p_\parallel$. Furthermore, one notices an additional caustic within the orbit 4 signal in panel (d) (starting at $\bm p \sim (1.9,0)$ a.u.) also carrying over to the orbit 3 signal. This caustic is similar to the 2-cycle caustic in panel (b), and is indeed linked to a structure similar to that in Fig.~\ref{fig:saddle_point_plot}(b1), but for the last parts of the 4-cycle pulse. 
The additional fine interference fringes found within the boundary of the caustic (compared to the 2-cycle case) thus represents interference between trajectories originating from two different cycles of the pulse. Other than these inter-cycle interference fringes, along with a shift of the caustic to higher $p_\perp$ values, the orbit 3 signal is not much affected by the increase in field cycles. 

Figure~\ref{fig:He_single_orbs} (c) and (f) shows the total signal from orbits 3 and 4, that is, the transition amplitude, Eq.~\eqref{CQSFA_amplitude_final}, for orbit 3 and 4 coherently summed before taking the norm square. For the two cycle pulse, panel (c), we only have a relatively small amount of interference between the two orbits, since the difference in signal of the orbit types is large (compare Fig.~\ref{fig:He_single_orbs} (a), (b)). The main interference effect is the emergence of fine interference fringes within the area of the left orbit 4 signal. For the 4 cycle pulse, Fig.~\ref{fig:He_single_orbs}(f), interference between orbit type 3 and 4 is more pronounced, since the difference in signal for the two orbit types is less. Especially, we see the emergence of the so-called spiral structure \cite{maxwell_spiral-like_2020} in the central parts of the PMD. The results here show that patterns coming from interference with orbit 4 depends most strongly on the few-cycle nature of the pulse, which was to be expected from the single orbit type PMDs. 

\subsection{Connections to initial momenta}

\begin{figure}
    \centering
    \includegraphics[width=\linewidth]{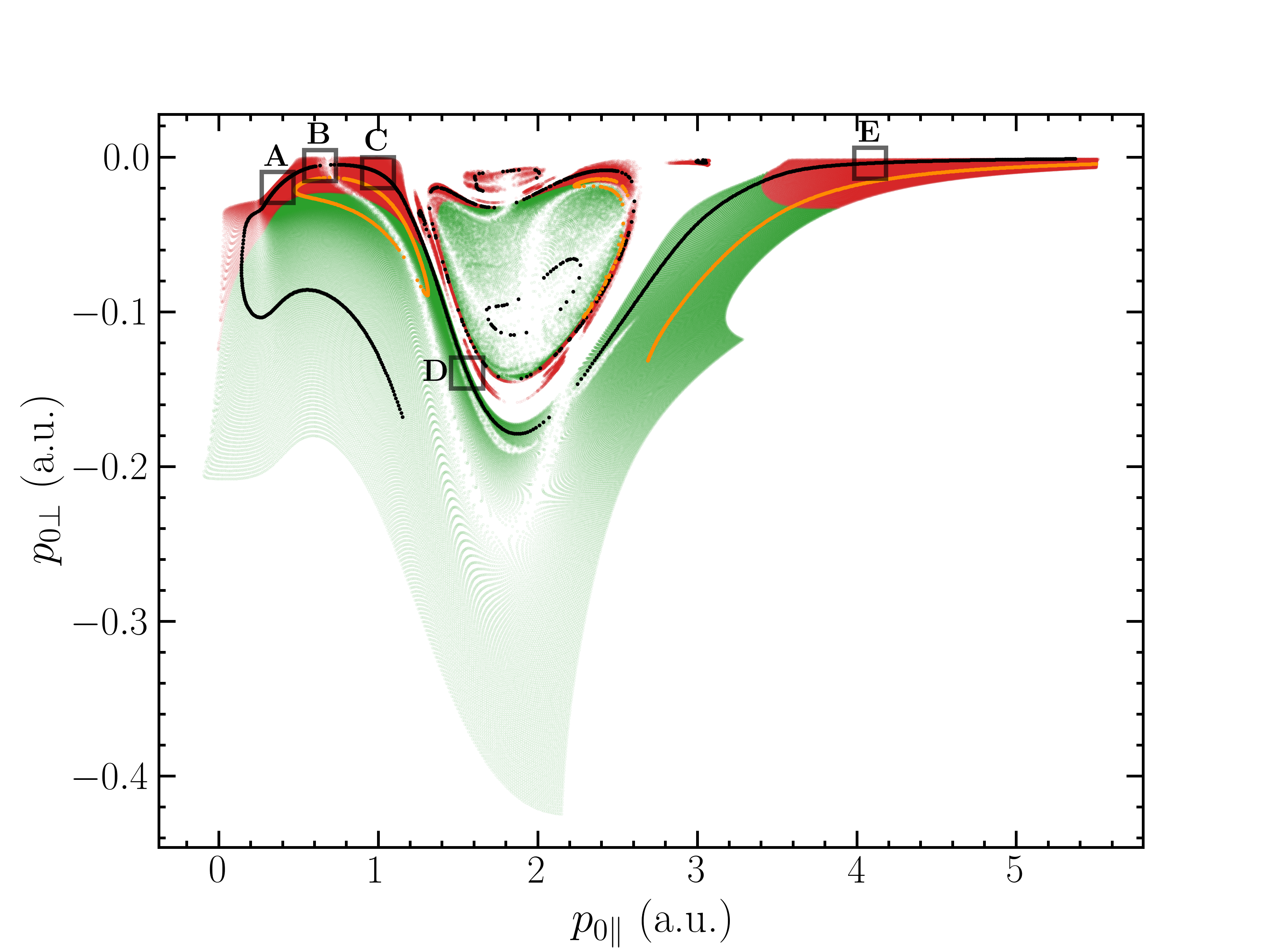}
    \caption{Initial momentum for orbit types 3 and 4 ionizing near the peak of the electric field for the He simulation of Fig.~\ref{fig:validation_PMDs} (structure in Fig.~\ref{fig:saddle_point_plot}(b1), but for all $p_\perp$). The green and red areas correspond to orbit 3 and 4 initial momenta, respectively. Black points corresponds to initial momenta for final perpendicular momenta of $p_\perp=0.25$ a.u.\ (Fig.~\ref{fig:saddle_point_plot}(b1)), while orange is for final momenta $p_\perp=1$ a.u. }
    \label{fig:init_orb34}
\end{figure}

The orbits found within the structure in Fig.~\ref{fig:saddle_point_plot}(b1) capture most of the orbit types 3 and 4 signal in the PMD. It is therefore interesting to investigate where in the initial momentum space these orbits originate from, and to optimize sampling of the initial conditions. In Fig.~\ref{fig:init_orb34}, we consider the initial momentum values for the orbits found in the central structure (Fig.~\ref{fig:saddle_point_plot}(b1))  across all values of final momenta. Furthermore, the specific initial conditions for the orbits shown in Fig.~\ref{fig:saddle_point_plot}(b1), with $p_\perp=0.25$ a.u., are marked with black dots. We also include the boxes A--E from Fig.~\ref{fig:saddle_point_plot}(b2), now mapped to initial momentum space. Clearly, judging by the density in the boxes in Fig.~\ref{fig:init_orb34}, the initial conditions in box A, B, C, and D, that are giving rise to the high signal structure in the PMDs Fig.~\ref{fig:He_single_orbs}, are located within a small area, $p_{0\parallel} \in [0;1.5]$ a.u., $p_{0\perp} \in [0;-0.2]$ a.u. The other parts of the figure, such as the orbits for $p_{0\parallel} > 2.5$ a.u., containing box E, were seen to have very little influence on the PMDs, even though they are densely populated in initial conditions. In addition, we have plotted the initial momentum values for the $p_\perp=1$ a.u.\ orbits, as seen in Fig.~\ref{fig:saddle_point_plot}(b1) in orange. Here, it is seen how the orbits giving the most signal in the PMDs (the ring structure close to box A, B, C) is even more localized in initial momentum space. 

Comparing with the total amount of sampled initial conditions, Fig.~\ref{fig:init_conds}, one sees that it is only a very small amount of the sampled trajectories that matters for the resulting PMDs. In particular, there are many densely populated areas of initial conditions that are low in probability, meaning a clustering approach to sampling quickly leads to unnecessary oversampling. For more complicated potentials, e.g., for molecular targets (and to some degree even for the Ar potential used here) this could make sufficient sampling very difficult. One way to counter this would be to sample in an iterative manner: First a set of random initial conditions are chosen, their trajectories propagated and the parameters important for their contribution to the transition amplitude, primarily $\Im{S(\tilde{\bm{p}}_s, \bm{r}_s, t_s)}$ and the Jacobian of the stability matrix $J_s(t)$, see Eq.~\eqref{CQSFA_amplitude_final}, are determined. Based on these parameters, one can estimate which regions of initial conditions should be sampled in greater detail, and prevent sampling in regions of low probability density. 

\section{Conclusion and outlook \label{sec:conclusion}}
In this paper we have presented an improved version of a semiclassical model for ATI, the CQSFA, and showcased its capabilities in reproducing results from the TDSE and ability to disentangle the intricate electronic dynamics on an attosecond timescale. To systematically verify the accuracy of the model we have analyzed the PMDs for three atomic targets, hydrogen, helium and argon, ionized by a two-cycle sin$^2$ pulse of linearly polarized light, and found excellent agreement. We have fully included the Maslov phase in the calculation of the transition amplitudes, which drastically improves the agreement with TDSE solutions. In earlier versions of the CQSFA, one found deviations in the holographic patterns in the PMDs, which have now been reconciled. Secondly, through a new, Monte-Carlo style sampling of the initial conditions, we solve the inversion problem by randomly sampling the initial momenta and propagating the trajectories forward in time. With this approach, we find `new' types of quantum trajectories, that were not previously included in the CQSFA. These make a new class of rescattered trajectories that go from laser-driven returns that ionize near the peak of the laser field (including the long and short solutions) to Coulomb-driven return that ionize near zero field.
Third, we have also expanded the model to accommodate few-cycle pulses of arbitrary shape, which is crucial for comparisons to state-of-the-art few-cycle ATI experiments, and solve in three dimensions.
%We have also shown the convenient ability to construct PMDs for a subset of the quantum orbits, which makes it possible to unravel the influences of each orbit type on the final PMD, or of any arbitrary combination of solutions to the SPEs. 

% Outlook-stuff? Bare slet det, hvis det er
The high level of accuracy of the CQSFA yields a new level of credibility to trajectory-based analysis within the field of strong-field physics. We have illustrated that it is possible to extract the trajectories most relevant for atomic ionization and relate them directly to interference structures in the PMD, enabling further optimization of the sampling algorithm.
This will enable the use of a trajectory-based analysis for more complicated targets, where the new trajectories highlighted could play a role. 
Furthermore, we imagine the CQSFA being used for strong-field simulations where direct TDSE solutions becomes unpractical, e.g., ionization of larger molecules, or longer wavelength, such as Ref.~\cite{maxwell_relativistic_2023}.
Access to accurate PMDs and trajectory analysis could give insight into attosecond dynamics, and even function as a sensitive chiral probe \cite{planas_ultrafast_2022}. 

\begin{acknowledgments}
ASM acknowledges funding support from the European Union’s Horizon 2020 research and innovation programme under the Marie Sk\l odowska-Curie grant agreement SSFI No.\ 887153. LBM acknowledges support from the Danish Council for Independent Research (Grant Nos.\ 9040-00001B and 1026-00040B).
\end{acknowledgments}

\appendix

\section{Further notes on the Maslov phase}\label{app:maslov}
In this appendix, we provide some additional information about the Maslov phase. More information about its properties and calculation may be found in Refs. \cite{levit_hamiltonian_1977, levit_focal_1978}. 

As mentioned in Sec. \ref{sec:theory}B, the calculation of the Maslov phase reduces to the estimation of the change in the Maslov index over the propagation of the classical trajectories arising from the SPA. To calculate the change in the Maslov index, one considers the second variation of the action within the Hamiltonian formalism. By doing so, one arrives at a boundary value problem, which may be transformed into a directly solvable Jacobi initial value problem. For a system with $n$ degrees of freedom, the latter takes the form 
\begin{equation}\label{Jacobi_ivp}
    \qty[\bm\Gamma \dv{\tau} - \bm\Theta(\tau)]\bm\xi^{(k)}(\tau) = \bm 0,\quad k=1,2,\ldots,n
\end{equation}
with the $2n\times 2n$ matrices $\bm\Gamma$ and $\bm\Theta$, with $\bm\Theta$ often referred to as the secondary Hamiltonian matrix. The matrices are defined as 
\begin{equation}
    \bm\Gamma = \mqty[\bm 0 & -\mathds{1} \\ \mathds{1} & \vb 0], \quad \bm\Theta = \mqty[\vb D & \vb C^\text{T} \\ \vb C & \vb B],
\end{equation}
where the $n\times n$ matrices $\vb B, \vb C$ and $\vb D$ are given by
\begin{equation}
\begin{gathered}
    B_{ij}(\tau) = \eval{\pdv[2]{H}{p_{s,i}}{p_{s,j}}}_{\tau}, \quad C_{ij}(\tau) =  \eval{\pdv[2]{H}{p_{s,i}}{r_{s,j}}}_{\tau}, \\ D_{ij}(\tau)= \eval{\pdv[2]{H}{r_{s,i}}{r_{s,j}}}_{\tau}.
\end{gathered}
\end{equation}
For the Hamiltonian considered in the main article, we have $\vb{C}=\bm0$. In the mixed representation, the $\bm\xi^{(k)}$s are of the form
\begin{equation}\label{xi_app}
    \bm\xi^{(k)}(\tau) = \dv{p_{0k}}\mqty[\bm p_s(\tau) \\ \bm r_s(\tau)],
\end{equation}
where $p_{0k}$ is the $k^\text{th}$ component of the initial momentum $\bm p_0$ at the start of the propagation, say, at $\tau=\Re(t_s)$. It is important to mention that Eq.~\eqref{xi_app} only holds within the mixed representation, and takes a different form in the momentum or position representation. Note that the first and last $n$ components of $\bm\xi^{(k)}$ represent the  $k^\text{th}$ column of the Jacobian fields $\partial\bm p_s(\tau)/\partial \bm p_0$ and $\partial\bm r_s(\tau)/\partial \bm p_0$, respectively, also known as the `stability matrices'. Their initial conditions are found in Eq.~\eqref{JIVP_init_conds}.

\bibliography{sovs}

\end{document}